# Hierarchy of neural organization in the embryonic spinal cord: Granger-causality graph analysis of *in vivo* calcium imaging data

Fabrizio De Vico Fallani[1,2,3,4,5], *IEEE member*, Martina Corazzol[2,4,*], Jenna R. Sternberg[1,2,3,4], Claire Wyart[1,2,3,4], Mario Chavez[2]

*Affiliations*:

[1] INSERM U1127, Paris, France ; [2] CNRS, UMR7225, Paris, France; [3] Sorbonne Université, UPMC Univ Paris 06, UMR S1127, Paris, France; [4] Institut du Cerveau et de la Moelle épinière (ICM), Paris, France; [5] Inria Paris-Rocquencourt, ARAMIS project-team, Paris, France; * *current address*: Centre de Neuroscience Cognitive, UMR 5229, CNRS, Bron, France

*Corresponding authors*: Fabrizio De Vico Fallani, fabrizio.devicofallani@gmail.com; Claire Wyart, claire.wyart@inserm.fr (for specific requests on zebrafish transgenic lines).

## Abstract

The recent development of genetically encoded calcium indicators enables monitoring *in vivo* the activity of neuronal populations. Most analysis of these calcium transients relies on linear regression analysis based on the sensory stimulus applied or the behavior observed. To estimate the basic properties of the functional neural circuitry, we propose a network-based approach based on calcium imaging recorded at single cell resolution. Differently from previous analysis based on cross-correlation, we used Granger-causality estimates to infer activity propagation between the activities of different neurons. The resulting functional networks were then modeled as directed graphs and characterized in terms of connectivity and node centralities. We applied our approach to calcium transients recorded at low frequency (4 Hz) in ventral neurons of the zebrafish spinal cord at the embryonic stage when spontaneous coiling of the tail occurs. Our analysis on population calcium imaging data revealed a strong ipsilateral connectivity and a characteristic hierarchical organization of the network hubs that supported established propagation of activity from rostral to caudal spinal cord. Our method could be used for detecting functional defects in neuronal circuitry during development and pathological conditions.

*Keywords*

Zebrafish, GCaMP3 Fluorescence, Functional connectivity, Neuronal networks, Graph modeling.

# 1 - INTRODUCTION

Neural circuits in the spinal cord play essential roles in vertebrate locomotion [1]. During early development, spontaneous coordinated activity rises from newborn neurons across the entire spinal cord. Embryonic and larval zebrafish provide a unique model to investigate the mechanisms involved in generating rhythmic patterns of neuronal activity due to their transparency, relatively small size and amenability to genetic manipulation [2]. Notably, the combination of genetics and optical imaging has been recently developed to investigate noninvasively the neuronal circuit function at single-cell resolution [3], [4]. In contrast to conventional electrophysiology, which records the activity of few neuronal cells through electrode insertion, calcium fluorescence imaging techniques allow to optically record from multiple neurons the Ca2+ influx associated with their activity and in systems as diverse as neuronal cultures, hippocampal brain slices, nematodes and zebrafish [5]–[13].

It has been reported that zebrafish embryos show spontaneous activity as early as 17 hours post-fertilization (hpf) [14]. This spontaneous activity consists in the contraction of the left and right trunk muscles alternatively without any external stimulus. This coiling behavior precedes touch-evoked escape response and swimming, which emerge in later developmental stages. Conventional electrophysiology analyses showed evidence in the paralyzed embryo for periodic and synchronized activity of motoneurons within one side of the spinal cord and alternation between the two hemicords [15]. Similar results were also replicated by using genetically encoded calcium indicators [16]; The latter approach, based on high frequency imaging (20Hz), has also shown rostro-caudal propagation of activity [16]. From a methodological perspective synchronized coordination between neurons has been generally evaluated by cross-correlation analyses, which are basic indexes of undirected functional connectivity [17]–[19]. Synchronization does not inform on the propagation of such activity, which could instead describe how the information flows across neuronal populations [20]. Current methods for investigating activity propagation in calcium fluorescence imaging data largely rely on the computation of either inter-spike delays between different signals [16], [17] or onset times of neuronal activation [10], [13], that can be seen as an heuristic instance of the more general cross-lagged correlation [21]. The arbitrary selection of several parameters (e.g. amplitude thresholds, refractory periods) to be fixed *a-priori*, as well as the sensitivity to spurious noisy peaks, make this heuristic method suboptimal in terms of general applicability. Interestingly, a recent *in vitro* study suggests that theoretically grounded methods, like the transfer entropy, can be effectively used to reconstruct activity propagation from calcium signals of neuronal cultures [12]. Despite its theoretical advantages, transfer entropy requires vey long data to obtain reliable estimates of activity propagation [22]. This technical requirement represents a limiting factor when dealing with *in vivo*, generally shorter, data.

In the present study, we use a more robust principled method to infer activity propagation by exploiting the entire dynamics of *in vivo* recorded calcium signals. Specifically we refer here to the Granger-causality (*GC*): a statistical concept, based on prediction error that is commonly used to quantify the information flow between neurophysiological signals [23]. Estimating Granger-causality between all the available calcium signals gives a functional connectivity pattern that describes in a compact way the directed interactions between all the neuronal cells. This functional connectivity pattern is shaped to some extent by the underlying synaptic connectivity [24] and provides complementary information about the putative dynamic flow within the neuronal network. Functional connectivity patterns can be probed as graphs or networks, i.e. mathematical objects composed of nodes corresponding to neurons and weighted directed links (*GC* values). Finally, we exploit graph theoretical approaches to *i)* characterize at a system level the putative functional circuitry between in the spinal cord of the zebrafish embryo and *ii)* identify the neuronal "hubs" of information flow that subserve the spontaneous activity.

## 2 - METHODS

### 2.1 EXPERIMENTAL DATA AND PREPROCESSING

All experiments were performed on Danio rerio embryos between 23 and 24 hours post fertilization. AB and TL strains of wild-type (WT) larvae were obtained from our laboratory stock of adults. Embryos and larvae were raised in an incubator at 28.5°C until 60-75% epiboly, then maintained at 26°C until shortly before recordings were performed. All procedures were approved by the Institutional Ethics Committee at the Institut du Cerveau et de la Moelle épinière (ICM). Five 28-30 somite-stage *Tg{s1020t:Gal4; UAS:GCaMP3}* [25]*)* transgenic embryos were used for this study. Before recordings were performed, embryos were dechorionated and screened for fluorescence. Embryos screened for GCaMP3 fluorescence were staged at 23 hours post fertilization (hpf) and embedded in 1.5% agarose with the dorsal side of the spinal cord oriented up. Embryos were anesthetized in MS-222 before being paralyzed with 0.5mM α-bungarotoxin (Sigma-Aldrich, USA) diluted in phenol red solution via injection of caudal muscle (segments 15-18). Data were acquired with a 3i spinning disk confocal microscope. Calcium imaging was performed during spontaneous activity for 250 seconds at 4 Hz with a 488 nm laser. A 20x objective allowed simultaneous imaging of five segments of the spinal cord from somites 3 to 7. After the experiment was performed, embryos were staged again to confirm that imaging reflected activity of 30-somite stage embryos (corresponding to 24 hpf). Images were acquired with Slidebook software and analyzed offline with ImageJ and MATLAB (The Mathworks, USA). Neurons expressing in these lines and implicated in the spontaneous activity are mainly motoneurons. They were identified based on the amplitude of the change of GCaMP3 fluorescence that is related to intracellular calcium concentration. ROIs for motoneurons on each side were manually selected based on the standard deviation image calculated over the entire recording. For each ROI, the fluorescence value was computed as *ΔF/F = (F(t)-$F_0$)/($F_0$-$F_{bg}$)*, where *F(t)* was the fluorescence signal averaged across the pixels within the ROI, the baseline fluorescence $F_0$ was calculated as the mean of the first 10 frames acquired and $F_{bg}$ was calculated in a region outside of the embryo (**Fig. 1**).

2.2 NEURONAL FUNCTIONAL CONNECTIVITY

To elucidate how different motoneurons were collectively recruited, we referred to the concept of functional connectivity [26]. Here, the organization of neural circuits is estimated on the basis of the model-free Granger-causality [27], which tests whether the prediction of the present value in one time series can be significantly improved by including information from another time series. If so, the second time series is said to have a casual effect on the first, and the degree of significance of the improvement may be taken as the strength of *GC*. The *GC* measure is typically implemented by autoregressive (AR) modeling. In AR modeling, a stationary signal $x_m(t)$ can be expressed by a linear regression of its past values according to the formula:

$$x_j(t) = \sum_{k=1}^{q} a_j(k)x_j(t-k) + e_1(t) \qquad 1)$$

where $a_j(k)$ are the regression coefficients of the univariate AR model, $q$ is the model order, and $e_1(t)$ is the respective prediction error. By introducing the information from a second stationary signal $x_i(t)$ the formula can be rewritten as:

$$x_j(t) = \sum_{k=1}^{q} a_j(k)x_j(t-k) + \sum_{k=1}^{q} a_i(k)x_i(t-k) + e_2(t) \qquad 2)$$

where $a_j(k)$ and $a_i(k)$ are the new regression coefficients of the bivariate AR model, and $e_2(t)$ is the new prediction error obtained by including also the past of $x_i(t)$ in the linear regression of $x_j(t)$.

The *GC* between $x_i(t)$ and $x_j(t)$ is then measured by the log ratio of the prediction error variances for the bivariate and univariate model:

$$GC_{i \to j} = ln\left(\frac{var[e_2(t)]}{var[e_1(t)]}\right) \qquad 3)$$

*GC* is a positive number; the higher $GC_{i \to j}$, the stronger the influence of $x_i$ on $x_j$ is. Such influence is often considered to reflect the existence of an information flow outgoing from the system *i* to the system *j* [28]. Finally, *GC* is generally asymmetric ($GC_{i \to j} \neq GC_{j \to i}$), which is an important feature to infer causal or diving relationships.

*2.2.1 APPLICATION TO CALCIUM FLUORESCENCE IMAGING DATA*

In our study the neuronal GCaMP3 fluorescence signals exhibited steady oscillations (see **Fig.1** for a representative embryo). Specifically, the 98.5% of the pooled data satisfied the stationary test [29] and we computed $GC_{i \to j}$ values, with $i,j = 1 \ldots N$ where *N* is the number of the identified motoneurons in each zebrafish embryo. The regression coefficients of the AR models were computed according to the ordinary-least-squares minimization of the Yule-walker equations. The model order *q* was selected for each zebrafish embryo according to the Akaike criterion [30]. This criterion tries to find the optimal q that minimizes the following cost function $IC(q) = \ln(det(\Sigma_2)) + \frac{8q}{T}$, where $\Sigma_2$ is the noise covariance matrix of the bivariate AR model and *T* is the number of samples of the time series. Basically, this cost function balances the variance accounted for by the AR model against the number of coefficients to be estimated. Obtained values across zebrafish embryos ranged between *8* and *20*, corresponding to sliding regression windows of *2* and *5 s*, a temporal period which is compatible with the dynamics of single calcium transients (see **Fig. 1**). Each zebrafish embryo was characterized by a full connectivity pattern by quantifying the *GC* influences between all the identified motoneurons in the spinal cord. The statistical significance of the obtained influences was established by means of a *F-test* under the null hypothesis $GC_{i \to j} = 0$ [23]. According to this procedure, only the *GC* values corresponding to percentiles inferior to a statistical threshold of *α=0.01*, Bonferroni corrected for multiple comparisons, were retained.

2.3 GRAPH MODELING OF FUNCTIONAL CONNECTIVITY

The obtained functional connectivity patterns were then analyzed by means of a graph theoretical approach, which is a powerful tool to characterize the organization of connected systems [31]. According to this framework, functional connectivity patterns can be regarded as graphs (or networks) composed of nodes (motoneurons) and directed weighted links (pairwise *GC* values). The mathematical representation of such network is an asymmetric square weighted matrix *W* of size *N*, where *N* is the number of motoneurons. The generic weight *w(j,i)* corresponds to the magnitude of the information flow estimated between the motoneuron *i* and the motoneuron *j*. Thus, $w(j,i) = GC_{i \to j}$ if the corresponding *GC* value is statistically significant and *w(j,i) = 0* otherwise.

*2.3.1 CONNECTIVITY MATRIX FACTORIZATION*

In general, functional networks describing neuronal systems are peculiar in that the nodes of the network represent neurons that are spatially embedded [32]. Human brain networks consist of two large subsets of neural assemblies coincident with the two hemispheres, this evidence playing an important role for the overall organization [33], [34]. Here, the obtained neuronal networks are embedded in a physical space coincident with the two sides of the spinal cord (i.e. hemicords) of the zebrafish embryo, where nodes are primary rhythmically active motoneurons [17]. Without loss of generality, we can assume that the first *M* nodes of the network belong to the left side (*L*) and the remaining *N-M* to the right side (*R*). The connectivity matrix *W* can be then usefully split in two subcomponents describing the ipsilateral and contralateral connectivity of the two hemicords:

$$W = W_{IL} + W_{IR} + W_C = W_I + W_C \qquad 4)$$

where $W_{IL}$ is a NxN matrix containing the first block MxM on the main diagonal of W and 0 elsewhere; $W_{IR}$ is a NxN matrix containing the second block *(N-M)* x *(N-M)* on the main diagonal of W and 0 elsewhere; $W_I$ is the sum of the blocks $W_{IL}$ and $W_{IR}$ corresponding to the ipsilateral connectivity of the hemicords; $W_C$ is a NxN matrix containing the two remaining blocks *(N-M)* x *M* and *M* x *(N-M)* on the second diagonal of W that correspond to the contralateral connectivity of the hemicords, and 0 elsewhere.

*2.3.2 CONNECTION INTENSITY*

We considered the ipsilateral and contralateral connectivity separately according to the factorization of *W*. The global intensity of connectivity was defined as the sum of all the weighted links within (ipsi) or between (contra) the two sides of the spinal cord and it was measured by the connection intensity *C*:

$$C_{ipsi} = \sum_{j \neq i=1}^{N} W_I(j,i), \quad C_{contra} = \sum_{j \neq i=1}^{N} W_C(j,i) \qquad 5)$$

By definition *C<sub>ipsi</sub>* and *C<sub>contra</sub>* are positive numbers; higher values of *C* reflect more intense connectivity. Connection intensity can be seen as the weighted version of the connection density for unweighted graphs [31].

*2.3.3 NODE STRENGTH*

At the local level we considered separately the connectivity of each node with respect to its ipsi/contralateral hemicord. Furthermore, we distinguished between incoming and outgoing connectivity. The local information flow was then measured by the node strength (or weighted degree), which considers the sum of all the incoming/outgoing weighted links:

$$D_{ipsi}^{in}(j) = \sum_{i=1}^{N} W_I(i,j), \quad D_{ipsi}^{out}(j) = \sum_{i=1}^{N} W_I(j,i) \qquad 6)$$

$$D_{contra}^{in}(j) = \sum_{i=1}^{N} W_C(i,j), \quad D_{contra}^{out}(j) = \sum_{i=1}^{N} W_C(j,i) \qquad 7)$$

By definition, node strengths are always positive and they are characterized by a very intuitive interpretation: nodes with high in-strength are highly influenced by others and represent potential receivers of information flow; nodes with high out-strength are central in the way they influence others and represent possible neural transmitter of information flow.

*2.3.4 Δ CENTRALITY*

We also defined a simple index that integrates the node strengths and gives a unique value of node centrality (or importance). This index of delta centrality $\Delta$ was computed as the difference between the out- and in-strength and it was defined either for the connectivity within (ipsi) and between (contra) hemicords:

$$\Delta_{ipsi}(j) = D_{ipsi}^{out}(j) - D_{ipsi}^{in}(j) \qquad 8)$$

$$\Delta_{contra}(j) = D_{contra}^{out}(j) - D_{contra}^{in}(j) \qquad 9)$$

$\Delta$ centrality values quantify the tendency of the nodes to act more as transmitter ($\Delta > 0$) or receiver ($\Delta < 0$) of information flow.

*2.3.5 NORMALIZATION BY RANDOM GRAPHS*

Due to experimental constraints (e.g. manual selection of motoneurons based on the variability of the fluorescence signal, see section 2.1), the resulting *GC* networks can have different sizes in terms of number of nodes and weighted links across embryos. The measured network indexes (i.e. connection intensity, node strength, delta centrality) were then normalized in each embryo to compensate possible biases on the network topology due to the different size [35]. Specifically, for each embryo, the measured network indexes were normalized by those obtained from a distribution of 100 equivalent graphs generated by randomly shuffling the weighted links of the actual *GC* network. From this distribution the mean and standard deviation were computed and used to normalize (Z-score) the actual network indexes. Eventually, a positive Z-score indicates that a network index measured on the actual *GC* network is higher than what expected in random graphs, a negative Z-score indicates the opposite, and a Z-score close to zero indicates that the network index measured on the actual *GC* network is similar to that obtained in equivalent random configurations.

# 3 - RESULTS

By performing calcium imaging within one plane of the ventral and rostral spinal cord, we observed that motoneurons of zebrafish 30-somite stage embryos generated large calcium transients reflecting synchronous bursting activity (see **Fig. 1**, for a representative embryo). Motoneurons were highly synchronized on each side, as illustrated in the cross-correlation connectivity matrix of **Fig 2a**. Reciprocal coordination is usually estimated by the standard cross-correlation coefficient (*CC*), which can only detect undirected relationships between neuronal activities, i.e. *w(j,i) = w(i,j)*. Here, we used Granger-causality measurements to infer the activity propagation between the calcium transients of the motoneurons. Such non-reciprocal relationships allow identifying a directed neuronal network as revealed by the asymmetric connectivity matrix *W* of the representative embryo (**Fig 2b**). In this matrix the elements above the main diagonal represent influences from rostral to caudal motoneurons, the elements below the main diagonal indicating the opposite direction. Since active neurons were selected based on the standard deviation of the fluorescence signal for the length of the recording, the number of considered motoneurons could differ across distinct embryos.

By using a graph theoretical approach, we found that the ipsilateral connectivity intensity $C_{ipsi}$ was significantly higher (*p<0.05*) compared with that obtained by randomly reshuffling the links (i.e. pairwise *GC*) among all the available motoneurons (here also referred as nodes). On the contrary, the contralateral connectivity intensity $C_{contra}$ was significantly lower (*p<0.05*) with respect to random graphs (**Tab. 1**). In general, $C_{ipsi}$ values tend to be higher than $C_{contra}$ (Wilcoxon Z-value=-1.75, p=0.079), this indicating that spontaneous coiling in the zebrafish was functionally characterized by a strong exchange of information between the nodes in the same hemicord.

At a local level, we characterized each node in terms of its outgoing/incoming information flow. The resulting information flows were preferably distributed along the same side of the spinal cord as compared to random graphs, i.e. $D_{ipsi}^{in} > D_{contra}^{in}$, $D_{ipsi}^{out} > D_{contra}^{out}$ (**Supplementary Fig. 1**). Given the scarce level of contralateral connectivity, both at the global and local scale, the ipsilateral graph metrics were expected to give more relevant information about the network organization. In particular, by fusing the information from the ipsilateral out-strength $D_{ipsi}^{out}$ and in-strength $D_{ipsi}^{in}$ we defined a compact delta centrality index ($\Delta_{ipsi}$) to identify the nodes acting as connectivity "hubs", i.e. nodes with relatively high information flow. The characteristic spatial organization of the $\Delta_{ipsi}$ values shows that rostral nodes tend to have a larger amount of outgoing information flow while caudal nodes have a higher tendency to receive information flow (see **Fig. 3**, for the representative embryo). By exploiting the actual longitudinal position of the motoneurons as measured by their Cartesian Y-coordinate from the field of view of the spinal cord, we tested whether the node $\Delta_{ipsi}$ centrality values and the normalized Y-coordinates of the corresponding motoneurons were statistically correlated. A first analysis considering the nodes in both the hemicords revealed a significant positive correlation for the pooled data (Spearman R=0.4157, p=0.0006 data not shown here). An even higher significant correlation (R=0.6001, p=0.0003) was obtained when we used the representative side of the spinal cord that gave the signals with better quality in each embryo (**Fig4a**). Given that the calcium imaging data are obtained by optical sectioning on a single plane across the spinal cord, one side of the cord often shows signals with higher amplitude and signal-to-noise ratio than the other. This is a well-known issue due to the difficulty to obtain in the same focal plane $\Delta F/F$ signals with exactly the same quality from neurons on both the hemicords (see for example **Fig. 1b**, motoneurons 1 and 2).

Such node spatial hierarchy could be also revealed, albeit marginally, by using node in-strength and out-strength values (**Supplementary Fig. 2**). The observed node hierarchy reflecting the rostro-caudal propagation of activity can be attributed to the superiority of the *GC* method over standard techniques using for instance lagged correlation (*LC*) to infer directionality [21]. To demonstrate this, the $\Delta$ centralities were computed for the *LC* networks obtained in each zebrafish embryo (**Supplementary Fig. 3**) and then correlated with the longitudinal position of the motoneurons. No significant Spearman correlation was reported when considering either the sides of the spinal cord (R=-0.089, p=0.482 data not shown here) or the representative side (R=0.037, p=0.837) (**Fig 4b**).

# 4 - DISCUSSION

Functional connectivity is considered fundamental to understand how neural assemblies exchange information and how the resulting network leads to physiological and pathological behavior [36]. Despite the high potential of connectivity-based approach in humans, neuroimaging techniques such as electroencephalography (EEG), magnetoencephalography (MEG) or functional magnetic resonance imaging (fMRI), only allow sampling of activity across a fraction of the entire neuronal network. This limitation restrains the power of subsequent analysis and results interpretation.

The transparency and small size of the zebrafish nervous system provide an opportunity to optically record calcium transients reflecting neuronal activity from the entire neuronal network [9], [16], [17], [37], [38]. Recently, the combination of optical and genetic techniques has allowed to record networks composed of thousands of neurons at single-cell resolution level [3], [4], [18], [19], [39], [40]. In the zebrafish embryonic spinal cord we show here that Granger-causality graph analysis of calcium imaging data, obtained from a single plane at the single cell resolution and low acquisition rate, can resolve the well-known rostro-caudal propagation in the spinal cord where cross-correlation analysis did fail. These findings are in line with previous studies showing that motoneurons are coupled via GAP junctions at this stage and that excitation most likely comes from descending fibers from the hindbrain [15], [41].

Applied to Z-scan measurements, our methodological approach could provide new insights on mechanisms underlying activity propagation and information flows between larger populations of neurons in dataset acquired on the whole brain [4], [18], [19], [39], [40], [42] and at faster acquisition rates [43]. However, when considering a large number of neurons ($N >> 10$), particular attention should be paid to the statistical validation of the estimated *GC* networks. In our study, the statistical threshold that filters out the less-significant links (section 2.2.1, Bonferroni criterion for multiple comparisons) inversely depends on the total number of possible of links $L=N*(N-1)$, i.e. the higher *L*, the more stringent the threshold is. This choice could lead to very sparse *GC* networks when $N >> 10$ and less stringent statistical procedures, based for example on false discovery rates [44], can be considered to retain a sufficiently higher number of weighted links describing the functional network.

Here, the resulting *GC* networks were factorized in order to disentangle the role of the connectivity within (ipsi) and between (contra) the two sides of the spinal cord. Our results suggest that the neuronal activity underlying the rhythmic dynamics at this early embryonic stage entails functional directed networks that foster the information propagation between ipsilateral neurons while discouraging contralateral connectivity flows. This outcome is in line with the previous findings obtained with undirected connectivity analyses, which reported high ipsilateral activity coordination and low contralateral correlation [10], [11]. The directed connectivity approach further revealed a characteristic spatial hierarchy of the network hubs, i.e. nodes with a relatively high incoming/outgoing information flow. Network hubs have a crucial structural role in connected systems as they convey the large part of the available information and mediate such information between other nodes [45]. The obtained results show that there are two classes of network hubs in the spinal cord of the zebrafish that underlie the early spontaneous activity: *i)* the *transmitter hubs*, i.e. nodes with a relatively high rate of outgoing information flow, and *ii)* the *receiver hubs*, i.e. those with a relatively high rate of incoming information flow. Interestingly, the transmitter hubs are located in the rostral zone of the spinal cord while the receiver hubs are rather located in the more caudal site (**Fig. 3, 4**).

Although our approach is limited in that it cannot directly unveil the underlying basic mechanisms giving rise to the measured *GC* networks, the observed characteristic network hub organization supports previous evidence demonstrating rostro-caudal activity propagation in terms of inter-spike delay between neurons [1], [16]. In those studies, rostro-caudal activity propagation had been observed with electrophysiology techniques and higher acquisition rate. Here, we showed that such hub organization cannot be determined when using temporal delays (i.e. cross-lagged correlation) between calcium imaging signals recorded at a low acquisition rate (**Fig. 4b**). Hence, the proposed network approach describing the functional circuitry in terms of information propagation between multiple neural activities appears to be *i)* more general, i.e. Granger-causality considers prediction error between signals whereas cross-lagged correlation takes into account simple delay and *ii)* more robust, i.e. rostro-caudal activity propagation can be characterized in terms of connectivity at lower acquisition rates.

In this study, the information propagation is estimated by means of bivariate AR models of Granger-causality (see Methods 2.2). Although other methods based on information theory [46] or multivariate autoregressive models (MVAR) could be used to assess directionality [47], practical evidence shows that they require generally longer data and strong assumptions on the absence of latent variables that could introduce spurious connectivity [48], [49]. Using MVAR models when these assumptions are marginally satisfied (like in our case) could lead to unreliable connectivity patterns. For the sake of completeness we reported the obtained MVAR *GC* networks in the **Supplementary Fig. 3**. A statistical procedure similar to the bivariate case (i.e. statistical threshold of *α=0.01*, Bonferroni corrected for multiple comparisons) was used to retain the significant *GC* influences. For these networks no significant hub hierarchy distribution was reported when correlating $\Delta_{ipsi}$ centrality values with the normalized Y-coordinates of the respective motoneurons either in both the sides of the spinal cord (R = 0.107, p = 0.393 data not shown here) or the representative side (R = -0.019, p = 0.916 data not shown here).

## 5-CONCLUSIONS

We used Granger-causality and graph theory to characterize the organization of neural functional connectivity patterns in the spinal cord of zebrafish embryos during spontaneous motor-like activity at the 30-somite stage. The obtained results allowed us to identify, a characteristic network structure that supports ipsilateral neural connectivity and rostro-caudal activity propagation. This general framework can provide effective network-based biomarkers for functional circuitry alteration of the spinal cord during natural development as well as in induced or spontaneous pathological conditions.

## ACKNOWLEDGEMENTS

The research leading to these results has received funding from the program "Investissements d'avenir" ANR-10-IAIHU-06. Mario Chavez was partially supported by the EU-LASAGNE Project, Contract No.318132 (FET STREP). This work received support from the Paris School of Neuroscience (ENP), the Fondation Bettencourt Schueller (FBS), Mr Pierre Belle, the City of Paris Emergence program, the Atip/Avenir junior program from INSERM and CNRS, the Fyssen foundation, the International Reintegration Grant from Marie Curie Actions Framework Program 6, and the European Research Council (ERC) starter grant "OptoLoco".

**FIGURE AND TABLE CAPTIONS**

**Figure 1. Optical imaging of calcium transients recorded in a 30-somite stage representative embryo in the stable transgenic zebrafish line *Tg(s1020t:Gal4; UAS:GCaMP3)*.** **Panel a)** Dorsal view showing the field of view. Scale bar = 50 μm. Rostral is up, "L" for left and "R" for right side of the spinal cord. Each motoneuron selected for analysis is circled (from 1 to 6 on the left side, and 7 to 11 on the right side). **Panel b)** ΔF/F fluorescence signals (related to calcium activity) for each motoneuron selected in a). Motoneurons 1 and 2 have a smaller ΔF/F signal-to-noise ratio, which may be due to their position slightly shifted compared to the focal plan. **Panel c)** The inset shows a zoomed view of a representative calcium transient for the motoneurons in the right hemicord.

**Figure 2. Cross-correlation (CC) versus Granger-causality (GC) connectivity matrix *W* of the representative zebrafish embryo.** The functional connectivity between motoneurons is coded by the elements of the *W* matrices. In these matrices, each element $w_{ij}$ contains the magnitude of functional connectivity between the activity of the neuron *i* and *j* according to the respective color bar. Elements on the main diagonal, indicating self-connections, were excluded due to their trivial interpretation and coded by a blue color (i.e. absence of connectivity). The first block identified by the rows and columns from 1 to 6 represent the connectivity between the motoneurons in the left hemicord (L), while the second block of rows and columns from 7 to 11 represent the connectivity between the motoneurons in the right hemicord (R). The intensity of the pairwise connectivity is coded by the colorbar. **Panel a)** The CC matrix shows a strong undirected (the matrix is symmetric) connectivity between ipsilateral motoneurons. **Panel b)** The asymmetric GC matrix shows a general tendency of the rostral motoneurons (lower identifier numbers) in each hemicord to influence caudal motoneurons (higher identifier numbers). Note that motoneurons 1 and 2 had a smaller ΔF/F signal-to-noise ratio (see **Fig. 1b**) and they are less involved in the connectivity matrices, i.e. low CC/GC values in the first two rows and columns of *W*.

**Figure 3. Rostro-caudal distribution of the nodal delta centrality in the representative zebrafish embryo. Panel a)** The normalized $\Delta_{ipsi}$ value is represented for each node (motoneuron) as a colored circle superimposed on the field of view. The larger the circle, the more central is the node in terms of its tendency to act as a transmitter (red color, positive value) or receiver (blue color, negative value) hub of information flow. **Panel b)** The same normalized $\Delta_{ipsi}$ centrality values are here represented within the neuronal *GC* network. Statistically significant *GC* influences are illustrated as directed arrows. The thicker the arrow the stronger the *GC* value is. Inter-hemicord directed links are illustrated in gray color for the sake of simplicity.

**Figure 4. Spearman correlation between nodal delta centrality and longitudinal position of the motoneurons for the pooled data.** Each circle in the panels represents the normalized Y-coordinate (x-axis, whereby positive values correspond to more rostral positions and negative values indicate more caudal positions) and the $\Delta_{ipsi}$ value (y-axis) for a node of a zebrafish embryo's neuronal network. The colors of the circles code the identifier of the embryo according to the legend. **Panel a)** shows a significant positive correlation for the neuronal networks obtained by computing Granger-causality between the calcium imaging signals. **Panel b)** no significant correlation was reported for the network obtained by using lagged correlation.

**Table 1. Normalized values of ipsilateral and contralateral connection intensity for the neuronal Granger-causality networks.** Asterisks denote significant deviations (abs(Z-score)>1.96, $p<0.05$) from equivalent random graphs. Group-averaged values are also reported at the end of the table.

# FIGURES

## Figure 1

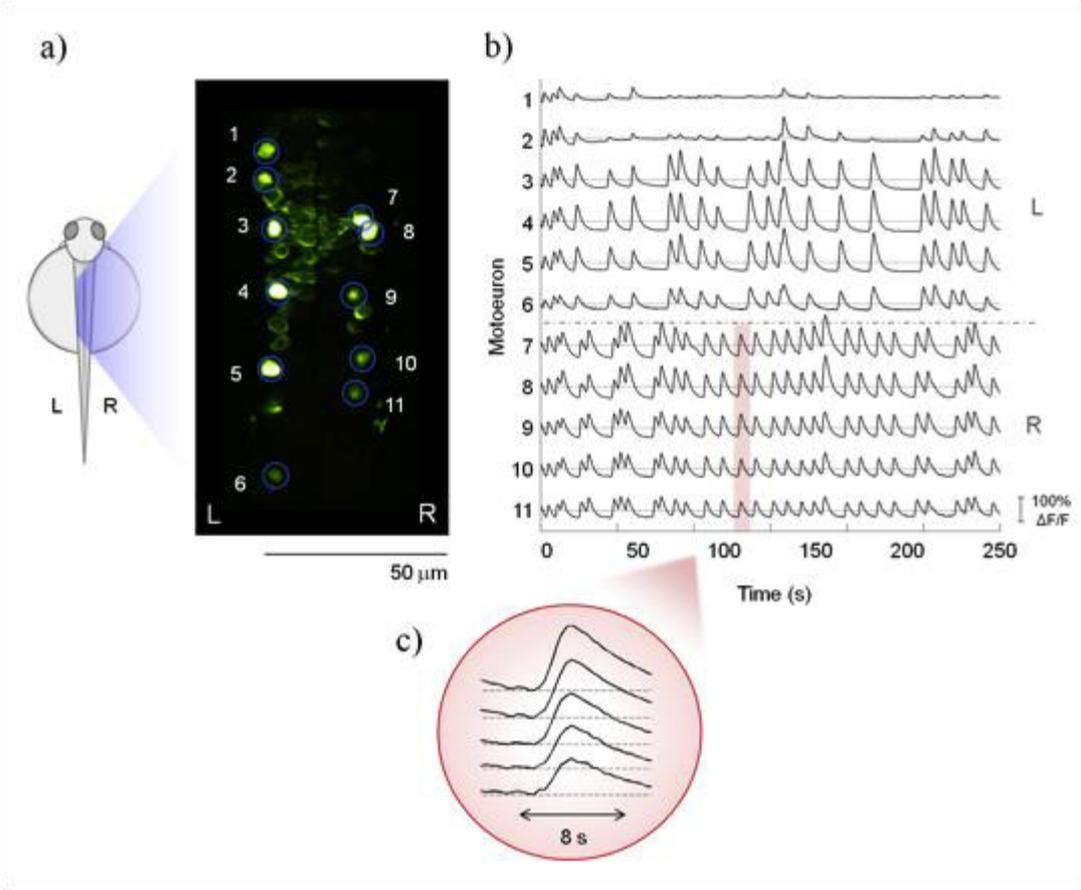

## Figure 2

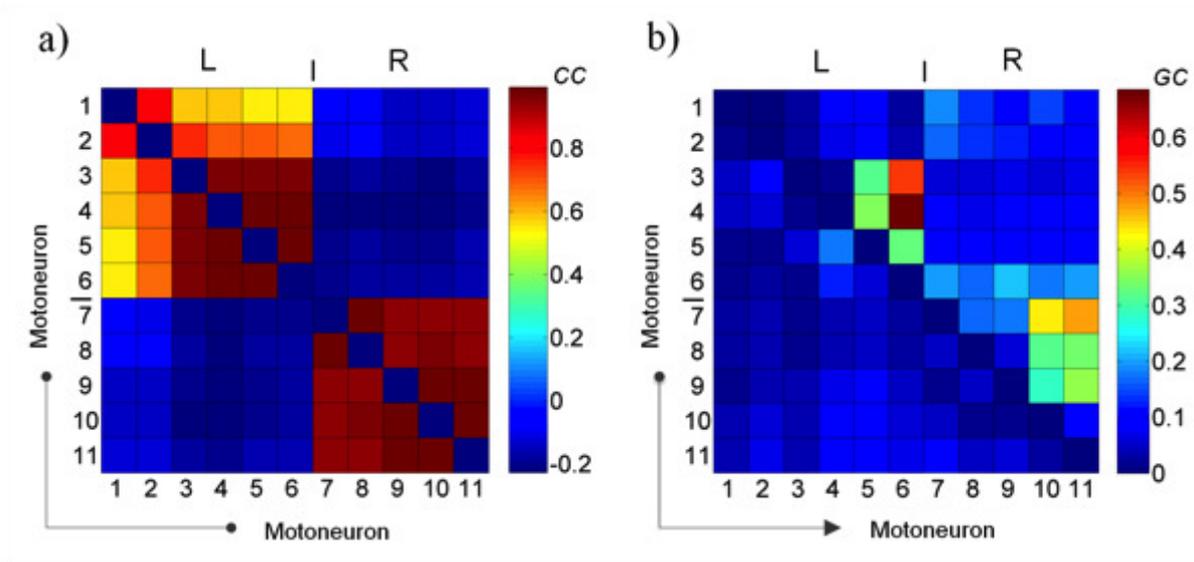

**Figure 3**

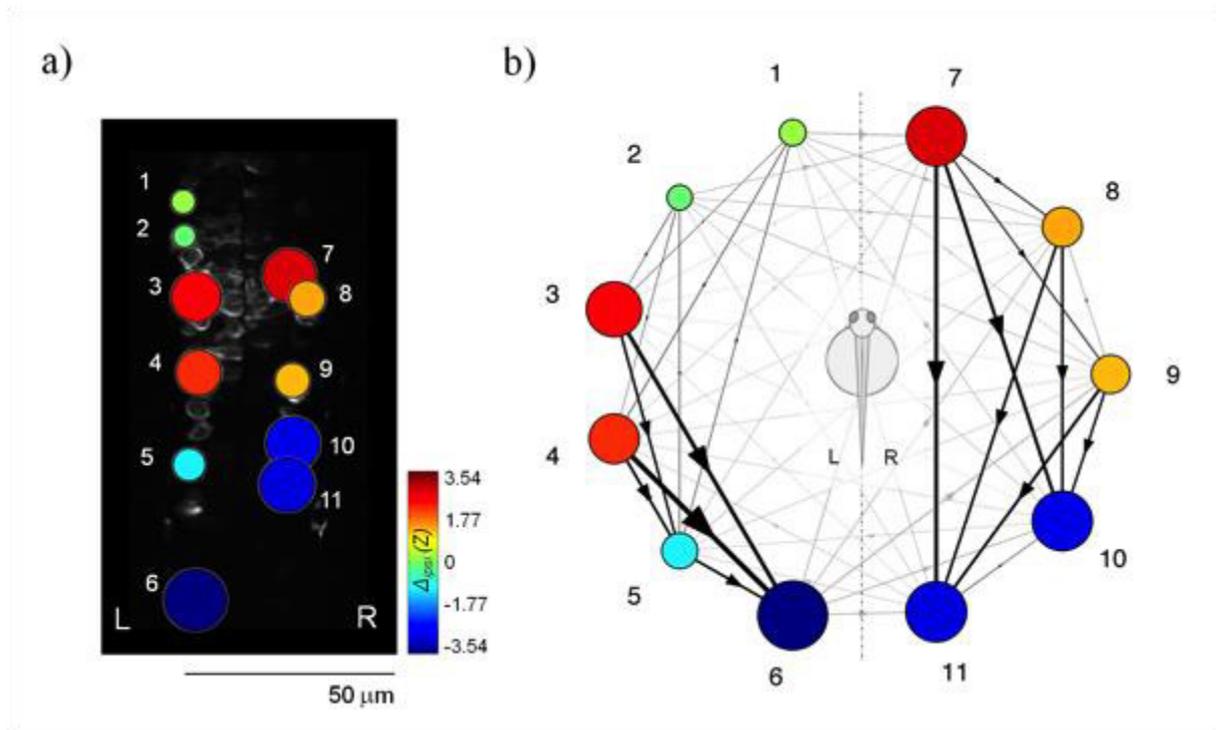

**Figure 4**

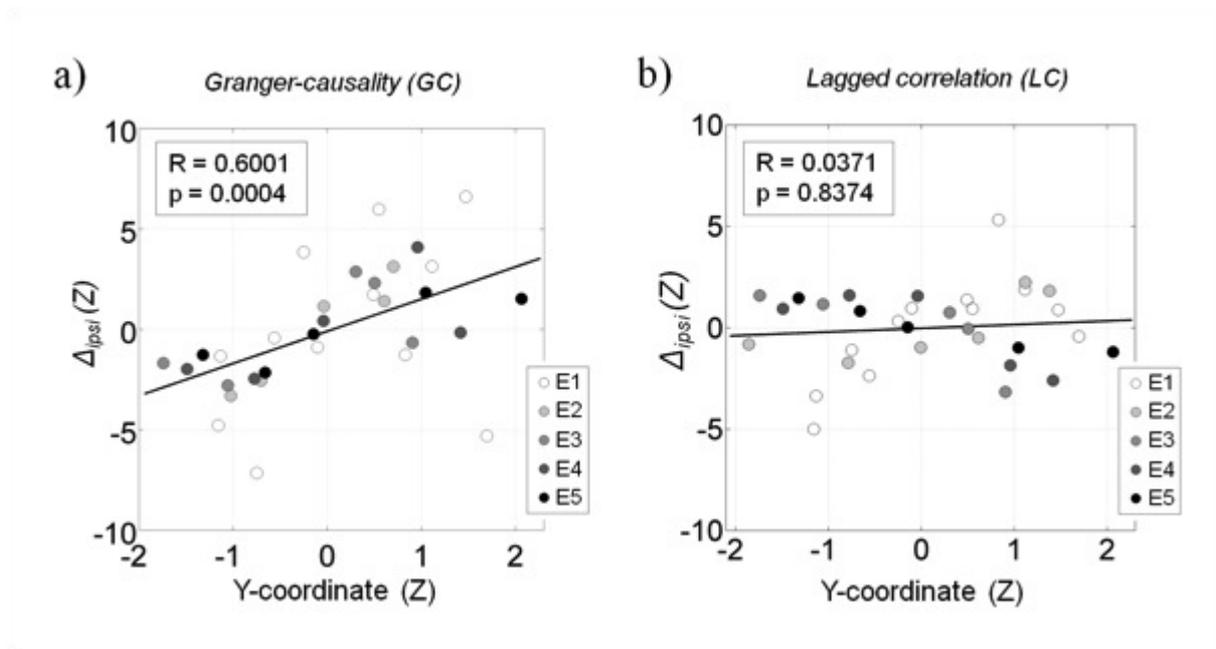

# TABLES

**Table 1**

|  | GC networks | |
|---|---|---|
| Embryo # | $C_{ipsi}$ (Z-score) | $C_{contra}$ (Z-score) |
| *E1* | 4.412* | -9.124* |
| *E2* | 1.982 | -4.436* |
| *E3* | 2.641* | -5.183* |
| *E4* | 2.653* | -5.203* |
| *E5* | 1.408 | -2.200* |
| **Group-average** | **2.619** | **-5.229** |